\documentclass[final,3p,times,twocolumn]{elsarticle}

\usepackage{lineno,hyperref}
\usepackage[utf8]{inputenc}
\modulolinenumbers[5]
\usepackage{graphicx}
\usepackage{comment}
\usepackage{subfig}
\usepackage{epsfig} 
\usepackage{caption} 

\journal{Nucl. Instr. and Methods in Phys. Res. Sec. B}

\begin{document}

 \captionsetup[figure]{labelfont={bf},name={Fig.},labelsep=period}
\begin{frontmatter}

\title{Particle Identification at VAMOS++ with Machine Learning Techniques}

\author[SNU,SNUPhys,cens]{Y.~Cho}
\author[cens]{Y.~H.~Kim\corref{cor1}}
\cortext[cor1]{Corresponding author}
\ead{yunghee.kim@ibs.re.kr}
\author[SNU,SNUPhys]{S.~Choi}
\author[cens]{J.~Park}
\author[cens]{S.~Bae}
\author[cens]{K.~I.~Hahn}
\author[SNU,SNUPhys,cens]{Y.~Son}
\author[GANIL]{A.~Navin}
\author[GANIL]{A.~Lemasson}
\author[GANIL]{M.~Rejmund}
\author[GANIL]{D.~Ramos}
\author[GANIL]{D.~Ackermann}
\author[GANIL]{A.~Utepov}
\author[GANIL]{C.~Fourgeres}
\author[GANIL]{J. C.~Thomas}
\author[GANIL]{J.~Goupil}
\author[GANIL]{G.~Fremont}
\author[GANIL]{G.~de France}
\author[KEK]{Y.~X.~Watanabe}
\author[KEK]{Y.~Hirayama}
\author[KEK]{S.~Jeong}
\author[KEK]{T.~Niwase}
\author[KEK]{H.~Miyatake}
\author[RIKEN]{P.~Schury}
\author[RIKEN]{M.~Rosenbusch}
\author[SKKU]{K.~Chae}
\author[SKKU]{C.~Kim}
\author[SKKU]{S.~Kim}
\author[SKKU]{G. M. ~Gu}
\author[SKKU]{M. J. ~Kim}
\author[Darmstadt]{P.~John}
\author[York]{A.~N.~Andreyev}
\author[CEA]{W.~Korten}
\author[Padova]{F.~Recchia}
\author[INFN_LNL]{G.~de Angelis}
\author[INFN_LNL]{R.~M.~P\'erez Vidal}
\author[INFN_Padova]{K.~Rezynkina}
\author[KULeuven]{J.~Ha}
\author[IPHC]{F.~Didierjean}
\author[CENBG]{P.~Marini}
\author[CENBG]{D.~Treasa}
\author[CENBG]{I.~Tsekhanovich}
\author[Lyon]{J.~Dudouet}
\author[VECC]{S.~Bhattacharyya}
\author[VECC]{G.~Mukherjee}
\author[VECC]{R.~Banik}
\author[VECC]{S.~Bhattacharya}
\author[RIKEN]{M.~Mukai}

\address[SNU]{Department of Physics and Astronomy, Seoul National University, Seoul 08826, Republic of Korea}
\address[SNUPhys]{Institute for Nuclear and Particle Astrophysics, Seoul National University, Seoul 08826, Republic of Korea}
\address[cens]{Center for Exotic Nuclear Studies, Institute for Basic Science, Daejeon 34126, Republic of Korea}
\address[GANIL]{Grand Accélérateur National d'Ions Lourds (GANIL), CEA/DRF-CNRS/IN2P3, F-14076 CAEN Cedex 05, France}
\address[KEK]{Wako Nuclear Science Center, IPNS, High Energy Accelerator Research Organization (KEK), Wako, Saitama 351-0198, Japan}
\address[SKKU]{Department of Physics, Sungkyunkwan University, Suwon 16419, Korea}
\address[Darmstadt]{Technische Universität Darmstadt, Karolinenplatz 5, 64289 Darmstadt }
\address[York]{School of Physics, Engineering and Technology, University of York, Heslington, York, North Yorkshire YO10 5DD, United Kingdom}
\address[CEA]{French Alternative Energies and Atomic Energy Commission, 17 rue des Martyrs 38054 Grenoble Cedex 9 France}
\address[Padova]{University of Padua, Via 8 Febbraio, 1848, 2, 35122 Padua, Italy}
\address[INFN_LNL]{INFN Laboratori Nazionali di Legnaro, IT-35020 Legnaro, Italy}
\address[INFN_Padova]{INFN Sezione di Padova and Dipartimento di Fisica e Astronomia dell'Universita', I-35131 Padova, Italy}
\address[KULeuven]{Katholieke Universiteit Leuven, Oude Markt 13, 3000 Leuven, Belgium}
\address[IPHC]{Institut Pluridisciplinaire Hubert Curien, Batiment 27, BP28, 67037 Cedex 2, 23 Rue du Loess, 67200 Strasbourg, France}
\address[CENBG]{Centre Etudes Nucléaires de Bordeaux Gradignan, 19 Chem. du Solarium, 33170 Gradignan, France}
\address[Lyon]{Insitut de Physique Nucléaire de Lyon, Université de Lyon, Université Lyon 1, CNRS-IN2P3, F-69622 Villeurbanne, France}
\address[VECC]{Variable Energy Cyclotron Centre, 1/AF, Bidhannagar, Kolkata, West Bengal 700064, India}
\address[RIKEN]{RIKEN Nishina Center, 2-1 Hirosawa, Wako, Saitama 351-0198, Japan}


\begin{abstract}
Multi-nucleon transfer reaction between $^{136}$Xe beam and $^{198}$Pt target was performed using the VAMOS++ spectrometer at GANIL to study the structure of n-rich nuclei around N=126.
Unambiguous charge state identification was obtained by combining two supervised machine learning methods, deep neural network (DNN) and positional correction using a gradient-boosting decision tree (GBDT). The new method reduced the complexity of the kinetic energy calibration and outperformed the conventional method improving the charge state resolution by $8$\%. 
\end{abstract}


\end{frontmatter}


\section{Introduction}

Multi-nucleon transfer (MNT) reactions near the Coulomb barrier gained renewed interest in accessing nuclides which are challenging to produce using conventional reactions~\cite{Adamian2020,Watanabe2015}.
One of the main challenges in such experiments using a magnetic spectrometer is particle identification because of the reaction fragment's large mass number (A), atomic number (Z), and wide range of charge states (Q). 

Recently, machine learning methods are being applied in nuclear physics theories and experiments, due to their effectiveness and ease of use.
Some applications of machine learning methods in particle identification are discussed
\cite{MLinNP2022}.

However, the applications of machine learning methods for obtaining continuous values such as particle energy (i.e. the regression problem) are not yet widely used in the analysis of experimental data \cite{MLinNP2022}. 

\section{Experimental setup}

The MNT reaction between $^{136}$Xe beam ($7$ MeV/u) and $^{198}$Pt target ($1.3$ mg/cm$^2$) was performed to populate neutron-rich nuclides towards the N=$126$ shell closure to investigate their nuclear structure. The experiment was carried out at GANIL using the large acceptance spectrometer VAMOS++ ~\cite{REJMUND2011, Kim2017} for detecting Projectile-Like Fragments (PLFs).

 The VAMOS++ particle detection system has two sets of multi-wire proportional counters (MWPC) near the target~\cite{VANDEBROUCK2016} and in the focal plane to measure the positions and angles of the particle for ion trajectory reconstruction and the time of flight (ToF) from the target to the focal plane for the velocity measurement. In addition, the ionization chamber (IC) with seven segments is located at the end of the focal plane part to measure the ion energy loss ($\Delta$E) and the total energy (E). 

\section{Paticle identification methods}
Particle identification of the PLFs at VAMOS++ is based on the combination of ToF, energy, and trajectory reconstruction on an event-by-event basis~\cite{REJMUND2011, PULLANHIOTAN2008}. Conventionally, the charge state (Q) was determined through energy measured by the IC ($E_{IC}$). $E_{IC}$ was calculated as the sum of energy loss ($IC_{i}$) in each segmented part of the IC ($E_{IC}=\Sigma a_{i}IC_{i}$), considering energy losses in the materials before the IC. Further corrections were made to account for the dependence of energy on other parameters. 

However, the conventional method had difficulty addressing the nonlinear energy loss behavior for low-energy ions before the IC. 
This made it difficult to use a single calibration parameter set over a wide energy range. 
To improve the accuracy of $E_{IC}$, a machine learning method was applied, providing greater selectivity in particle identification, 
especially for the rarely produced isotopes.

\subsection{Ion energy calculation using Deep Neural Network (DNN)}
Supervised Learning (SL) for regression is a framework of machine learning to find the best approximate function from the inputs to the targets, which are the correct answers\cite{ML_book}. We applied the SL framework to the present dataset using the ROOT TMVA module~\cite{Hocker2007}.

DNN \cite{goodfellow2016deep} is one of the SL's well-known tools, mimicking the neural network in the human brain. DNN consists of an input layer, several hidden layers, and an output layer with a number of nodes.

\begin{figure}[t]
\includegraphics[width=0.44\textwidth]{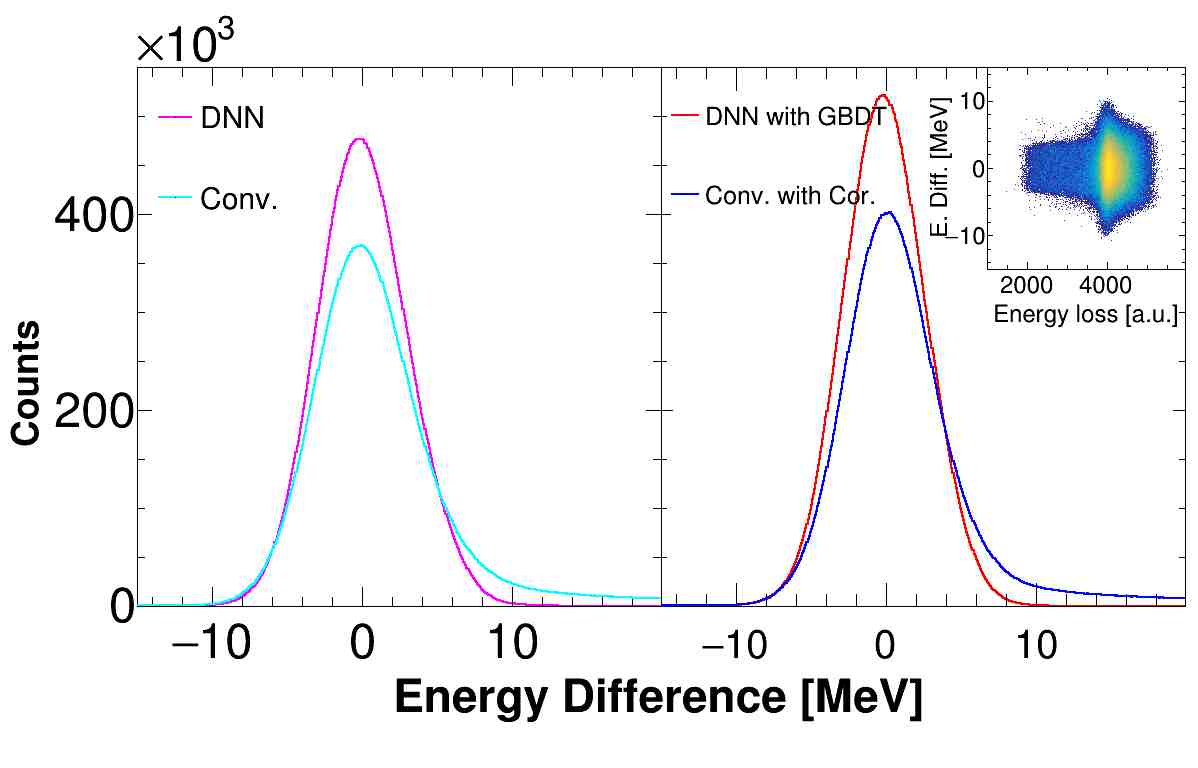}
\caption{The difference between the energy obtained from the reconstruction ($E_{REC}$) and the energy obtained from the ionization chamber using different methods. The cyan, magenta, blue, and red lines show the results of conventional method, DNN, conventional method with correction, and DNN with GBDT, respectively. The inset shows the energy difference of the DNN with the GBDT method versus the energy loss in the first segment of the ionization chamber.}
\label{fig:DNN_Ediff}
\end{figure}
A DNN with 7-10-3-1 layers  was used to calculate the kinetic energy of the PLF. The training data set comprised energy losses in the segmented IC as inputs and the $E_{REC}$ as the target. The first hidden layer used hyperbolic tangent (tanh) activation functions, while the second hidden layer used ReLU activation functions \cite{goodfellow2016deep}. Using the Adam optimizer~\cite{Adam2014}, the DNN was trained to minimize the mean square error between its output ($E_{DNN}$) and $E_{REC}$.

The validity of the output was checked by comparing $E_{DNN}$ and $E_{REC}$.
Fig. ~\ref{fig:DNN_Ediff} left presents the improvement of the ion energy calculation compared to the conventional method. The charge state could be calculated using $E_{DNN}$ after training the DNN. However, it was dependent on the entrance position of the IC (Fig. ~\ref{fig:QvsX}(a)) because of the bulged shape of the mylar window causing position dependencies in the energy loss, which needs an additional correction. 

\subsection{Correction of position dependence using Gradient Boosting Decision Tree (GBDT)}

The GBDT \cite{Friedman2001} is an algorithm to enhance the overall accuracy and stability of a single decision tree method \cite{Tree1984} by combining a series of trees. Each tree predicts the errors from the previous tree and the final output of the GBDT is the weighted sum of the trees' outputs. 

\begin{figure}[t]
\centering
\includegraphics[width=0.44\textwidth]{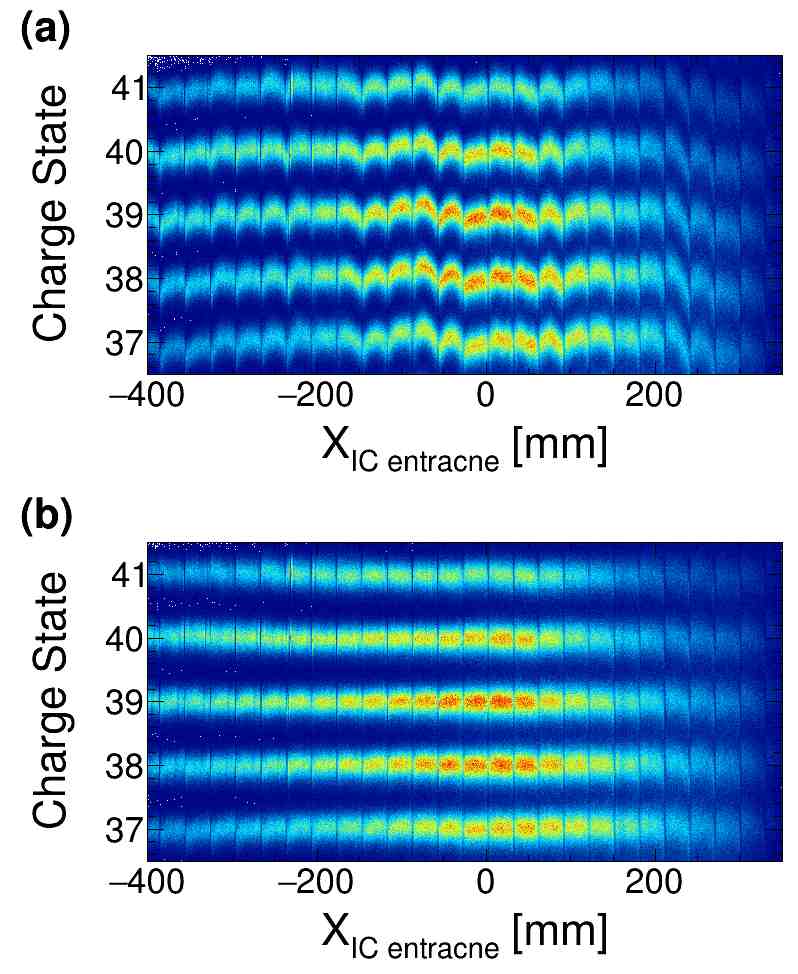}
\caption{2D spectra of the charge state versus the horizontal position at the entrance of the ionization chamber (a) before and (b) after the GBDT correction.}
\label{fig:QvsX}
\end{figure}

The GBDT was used to correct for the minor effects on the kinetic energy by calculating the residual energy between the $E_{REC}$ and the $E_{DNN}$. The inputs of the training data set were horizontal and vertical positions at the entrance of the IC. 
The target was residual energy divided by the energy loss in the first segment of the IC. Five hundread trees were created with a maximum depth of 4 optimized using the Huber loss~\cite{Huber1964}.

\begin{figure}[t]
\includegraphics[width=0.44\textwidth]{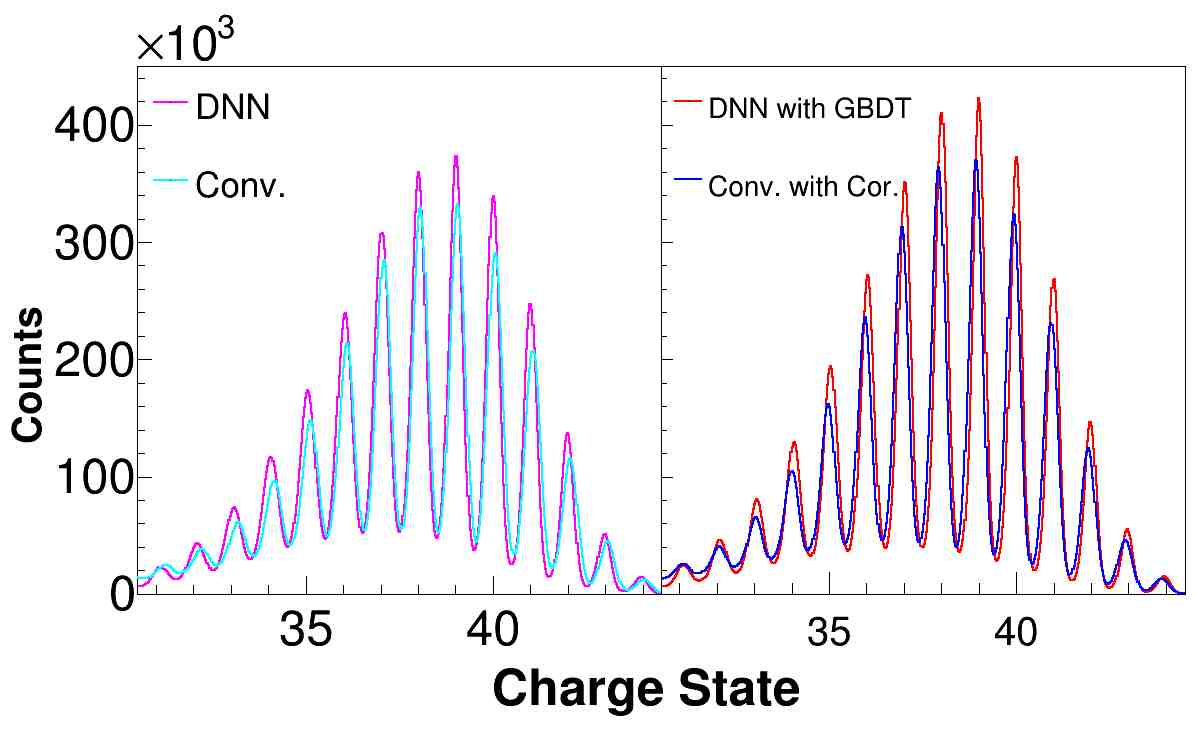}
\caption{The charge state distribution obtained using different methods. The cyan, magenta, blue and red line shows the result of the conventional method, DNN, conventional method with correction and DNN with GBDT respectively}
\label{fig:Qprojection}
\end{figure}

 The red line on the right of Fig.~\ref{fig:DNN_Ediff} illustrates that GBDT combined with DNN outperformed the conventional method in calculating the ion energy. The inset shows the residual energy versus the energy loss at the first segment of the IC for the case of GBDT combined with DNN, demonstrating the good training result over the whole energy loss range. Fig. ~\ref{fig:QvsX}(b) indicates that the GBDT accounts for the horizontal position dependence at the IC entrance for a charge state. Other minor dependencies such as vertical position dependence were also minimized.
Fig. ~\ref{fig:Qprojection} compares the charge state calculated by the conventional method, DNN, and the combination of DNN with GBDT. 
The charge states were better aligned at integer values, and the resolution ($\Delta$Q/Q) of DNN (conventional method) was $1/78$ ($1/70$) and of DNN with GBDT (conventional method with correction) was $1/86$ ($1/79$), which is a significant improvement due to a more accurate energy derivation.

\section{Summary}
A new calibration method of particle identification for the VAMOS++ spectrometer, utilizing supervised machine learning, was developed for the first time to aid unambiguous particle identification for rarely produced nuclides near N=126 shell closure.
A DNN and GBDT algorithm was used to determine the ion energy from the segmented IC.
The obtained energy showed good agreement with the reconstruction-based energy within 6.52 MeV, automatically correcting different minor effects. 
The charge state resolution was improved by $8$\% demonstrating the effectiveness of the new method. 

\section*{Acknowledgement}
The authors thank for the support of the National Research Foundation of Korea (NRF) grants funded by the Korean government (MSIT), Grants No. 2022R1A2C2005093, No. 2019K2A9A2A10017576 and No. 2020R1A2C1005981;
the Institute for Basic Science, South Korea (IBS), Grants No. IBS-R031-D1;
European Union’s Horizon 2020 research and innovation programme under grant agreement No. 654002. The data presented here originates from the E806\_20 GANIL dataset ~\cite{e806DataSet}.  We express our gratitude for the support of the GANIL staff and the AGATA collaboration.


\bibliography{ref}

\end{document}